\documentclass[
    twocolumn, superscriptaddress, nofootinbib, amsmath, amssymb,aps,prd, floatfix
]{revtex4-2}
\usepackage[dvipsnames,table]{xcolor}
\usepackage{graphicx,epsfig,psfrag,bm,amssymb}
\usepackage[colorlinks=true,urlcolor=blue
,anchorcolor=blue,citecolor=blue,filecolor=blue,linkcolor=blue,menucolor=blue
]{hyperref}
\usepackage{dcolumn}
\usepackage{bm}
\usepackage{xspace}
\usepackage{color}
\usepackage{soul}
\usepackage{cancel}
\usepackage{mathrsfs,amsfonts,color}
\usepackage{caption}
\usepackage{subcaption}
\usepackage{comment}
\usepackage{xspace}
\usepackage[capitalise]{cleveref}
\usepackage{ulem}
\usepackage{cancel}
\usepackage[export]{adjustbox}
\usepackage{placeins}

\newcommand{\vecbf}[1]{{\bm{#1}}}
\newcommand{\beq}{\begin{equation}}
\newcommand{\eeq}{\end{equation}}

\newcommand{\GeV}{\xspace\text{GeV}}
\newcommand{\Ap}{A^\prime}

\DeclareMathOperator{\br}{BR}
\DeclareMathOperator{\KL}{KL}
\newcommand{\pythia}{\textsc{Pythia}~8.3\xspace}

\begin{document}
\title{Data-Driven Predictions for Dark Photon and Millicharged Particle Production}
\author{Elizabeth Allison}
\affiliation{Neuroscience Graduate Program, Schulich School of Medicine and Dentistry, Western University, London, Ontario, Canada}
\affiliation{Department of Physics and Astronomy, York University, Toronto, Ontario, Canada}
\author{Nikita Blinov}
\email{nblinov@yorku.ca}
\affiliation{Department of Physics and Astronomy, York University, Toronto, Ontario, Canada}
\date{December 3, 2025}
\begin{abstract}
Accurate signal predictions are essential for interpreting and optimizing fixed-target searches for new physics. Even in minimal models such as the dark photon ($A'$) or millicharged particles (mCPs), theoretical uncertainties in hadronic production can be substantial. We introduce a data-driven framework that predicts both the rate and kinematic distributions of $A'$ and mCP production directly from measured dilepton events, without relying on specific theoretical production models. This method uses the close correspondence between amplitudes for emission of $A'$ or mCPs, and for off-shell Standard Model photon production, the latter being experimentally measurable in full differential form. We demonstrate that normalizing flow models can learn these distributions from data and serve as a fast, realistic Monte Carlo generator for dark sector signal simulations.
\end{abstract}
\maketitle

\section{Introduction}
Fixed-target experiments are a powerful way of searching for light, weakly coupled particles in a variety of dark sector models~\cite{Beacham:2019nyx,Batell:2022dpx}. Interpreting these searches, however, requires accurate predictions for both the production rate and kinematic distributions of potential signals. Even in minimal extensions of the Standard Model (SM), such as the dark photon ($\Ap$) or millicharged particles (mCPs), these predictions can carry significant theoretical uncertainty, particularly at proton beam experiments where hadronic production mechanisms are not well constrained. Such uncertainties propagate directly into signal acceptance estimates and ultimately affect the sensitivity reach of ongoing and future searches.

A key feature of $\Ap$ and mCP models is that their interaction with SM particles is through electromagnetism. As a result, their production amplitudes are closely related to the amplitude for emission of an off-shell SM photon. This insight was used by \cite{Ilten:2016tkc} to develop a data-driven dark photon search strategy at the LHCb experiment. Here we extend this idea to fixed-target experiments in the context of both dark photons and millicharges. We will show that fully differential distributions of virtual photons, which are accessible experimentally through dilepton measurements, encode precisely the information needed to predict the rate and kinematics of $\Ap$ and mCP production.

In order to make use of the theoretical connection between $\Ap$/mCP production and off-shell photon emission in a realistic setting, we will employ a machine learning approach to model the dilepton kinematic distributions. Normalizing flows (NFs) provide a flexible way to achieve this and they offer fast sampling from the resulting model. By training a conditional NF on dilepton data, we construct a fully differential, data-driven Monte Carlo (MC) generator that can be used to simulate both dark photon and millicharged particle signals. This allows fixed-target experiments to anchor their signal modeling directly to measured SM processes.

The rest of this paper is organized as follows. In \cref{sec:ap_production} we derive a simple relationship between the differential cross-sections for the production of $A'$ or mCPs and dileptons. This result is the foundation of our method, enabling us to relate the dark photon or mCP rates and kinematics to observables in \cref{sec:data_driven}. There we illustrate how to leverage this relationship in the context of a mock data set tuned to reproduce published results from the NA60 fixed-target experiment. We also describe a normalizing flow architecture that can be used to model the observed kinematic distribution of dilepton pairs. This produces a data-driven Monte Carlo generator for $A'$ or mCP kinematics. We discuss further applications of our approach and conclude in \cref{sec:conclusion}. \cref{sec:nf_training} contains details of training and hyperparameter studies for the normalizing flow models.

\section{Production Mechanisms\label{sec:ap_production}}
In order to effectively search for dark sector particles, one needs to identify the set of production channels relevant to each experiment, evaluate the kinematics of potential signal events, and compute the expected signal yield as a function of model parameters. At lepton beam experiments, dominant production modes of the dark photon include bremsstrahlung $\ell N \to \ell N + A'$, annihilation $e^+ e^- \to A'$ and inverse Compton-like processes $\gamma e^- \to e^- A'$~\cite{Bjorken:2009mm,Nardi:2018cxi,Marsicano:2018krp,Celentano:2020vtu}. The scattering matrix elements and kinematic distributions that follow from them are perturbatively calculable, even for bremsstrahlung which involves scattering off the nucleus -- the relevant hadronic form-factor is known. Millicharged particles are produced through the same electromagnetic processes: replacing the final-state $A'$ with $\bar{\chi}\chi$ yields the corresponding mCP channels, whose rates can be computed with similar theoretical precision.

Hadronic beam experiments introduce a qualitatively different challenge. Dark photons and mCPs can be produced through numerous channels, including rare meson decays, hadronic bremsstrahlung and mixing with vector mesons~\cite{Kyselov:2024dmi}. Some of these are straightforward to model, while others have significant uncertainties. Signal predictions typically rely on a combination of existing Monte Carlo generators, such as {\sc Pythia}~\cite{Bierlich:2022pfr}, or custom codes~\cite{deNiverville:2016rqh} designed for modelling low-energy hadronic interactions. These tools may not be validated in the low transverse momentum ($p_T$), forward region relevant for fixed-target searches. Additionally, some processes, such as coherent proton bremsstrahlung, are not included in most simulations and must be added by hand~\cite{Blumlein:2013cua,Foroughi-Abari:2021zbm, Gorbunov:2023jnx, Foroughi-Abari:2024xlj,Kling:2025udr}. These calculations contain hadronic inputs, i.e., form-factors, that are not currently well-constrained by data, and therefore lead to significant systematic uncertainties in signal predictions and data interpretation. Below we develop a method to calibrate dark photon and mCP production rates and kinematics \emph{in situ} using dilepton measurements in the relevant detector, thereby bypassing many of the difficult modelling problems described above.

We imagine that an experiment has measured the kinematic distribution of off-shell SM photons through observations of, e.g., $\gamma^* \to \ell^+ \ell^-$. We want to leverage this measurement to predict the distribution and yield of $\Ap$ or mCPs as a function of their coupling and mass. To do this, we will find a relationship between the $\gamma^*$ and $\Ap$ or mCP distributions. 

First, we consider generic production of $\gamma^* \to \ell^+ \ell^-$. At the relatively low center-of-mass energies of interest for fixed-target experiments (well below the electroweak scale), the amplitude can be written schematically as
\beq
\mathcal{M}_{\gamma^*} = 
\includegraphics[valign=c,raise=0.0ex]{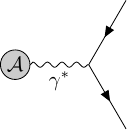} 
= \frac{e}{q^2} \mathcal{A}_\mu(q) \bar{u}_\ell \gamma^\mu v_{\bar\ell}
\eeq
where $\mathcal{A}_\mu$ is part of the amplitude that describes the production of $\gamma^*$ with (off-shell) momentum $q$; $u_\ell$ and $v_{\bar{\ell}}$ are Dirac spinors corresponding to the outgoing leptons. The initial and final state can involve other particles, which are not shown. Using this parametrization we can compute the distribution $d\sigma_\gamma$,  differential in $\gamma^*$ mass $m_{\gamma}^2 \equiv q^2 = (p_{\ell^+} +p_{\ell^-})^2 = m_{\ell\ell}^2$ and three-momentum $\vecbf{q}$, with the result
\begin{align}
\frac{d\sigma_{\gamma}}{dm_{\gamma}^2 d^3 q} 
& = \mathcal{F} \frac{\alpha}{48\pi^4 m_\gamma^2 q^0}\sqrt{1-\frac{4m_\ell^2}{m_\gamma^2}}\left(1 + \frac{2m_\ell^2}{m_\gamma^2}\right) \nonumber\\ 
& \times \int d\Phi \left[\frac{q^\mu q^\nu}{m_\gamma^2} - g^{\mu\nu}\right]\mathcal{A}_\mu\mathcal{A}^*_\nu
\label{eq:dilepton_differential_xsec}
\end{align}
where $\alpha=e^2/4\pi$, $\mathcal{F}$ is the usual cross-section flux factor (which depends on the specific initial state), $q^0 = \sqrt{m_\gamma^2 + \vecbf{q}^2}$ is the off-shell photon energy, and $\int d\Phi$ is the Lorentz-invariant phase space integral over any final state particles except for $\ell^\pm$. To obtain the above distribution we inserted 
\beq
1 = \int dm_\gamma^2 d^4 q \delta^4(q - p_\ell - p_{\bar{\ell}})\delta(m_\gamma^2 - q^2)\theta(q^0)
\eeq
and carried out the phase space integrals over $\ell^\pm$. We will perform similar calculations for the emission of $\Ap$ and mCPs, and relate the results to \cref{eq:dilepton_differential_xsec}.
\subsection{Dark Photon}
Dark photons interact with the SM through a kinetic mixing with the photon~\cite{Holdom:1985ag,delAguila:1988jz}. \footnote{We work far below the scale of electroweak symmetry breaking and do not consider the mixing with the hypercharge gauge boson.} Upon diagonalization, one finds that electromagnetic currents $J^\mu_{\mathrm{EM}}$ couple to the $\Ap$ just as they do to the photon, but with a strength suppressed by the dimensionless mixing parameter $\varepsilon$:
\beq
\mathcal{L}\supset e (A_\mu +  \varepsilon A'_\mu) J^\mu_{\mathrm{EM}}.
\label{eq:dark_photon_coupling}
\eeq
Using this interaction we compute the emission of a massive on-shell dark photon with polarization $\epsilon^\mu$:
\begin{equation}
\mathcal{M}_{\Ap} = 
%
\includegraphics[valign=c,raise=0.0ex]{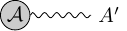} 
= \varepsilon \mathcal{A}_\mu(q) \epsilon^\mu(q)^*.
\label{eq:dark_photon_production_amp}
\end{equation}
Crucially, we assume that the amplitude $\mathcal{A}$ is the same as for $\gamma^*$; we expect this to be a sound approximation since weak neutral currents would lead to corrections only of $\mathcal{O}(m_{\Ap}^2 / m_Z^2)$.

Squaring \cref{eq:dark_photon_production_amp} and summing over $\Ap$ polarizations, we find a simple relationship between the dark photon differential cross-section $d\sigma_{A'}$ and $d\sigma_\gamma$:
\begin{align}
\frac{d\sigma_{\Ap}}{d^3 q} & = \varepsilon^2 \left[\frac{\alpha}{3\pi }\sqrt{1-\frac{4m_\ell^2}{m_{\Ap}^2}}\left(1 + \frac{2m_\ell^2}{m_{\Ap}^2}\right)\right]^{-1} \nonumber \\
& \times m_{\Ap}^2\left.\frac{d\sigma_\gamma}{dm_{\gamma}^2 d^3 q} \right|_{m_\gamma^2 = m_{\Ap}^2}
\label{eq:ap_from_gamma_differential_xsec}
\end{align}
An analogous formula holds for decay rates with the replacement $\sigma_{A'}\to\Gamma_{A'}$, $\sigma_\gamma \to \Gamma$, where $\Gamma$ is the decay width of a particle decaying into final states containing $\gamma^*$. 
For example, one can easily check that this relation holds between specific processes such as $\pi^0 \to \Ap \gamma$ and $\pi^0 \to \gamma(\gamma^*\to e^+e^-)$.  Similar formulae appear in hadronic bremsstrahlung literature, relating cross-sections for virtual photon emission and dilepton production~\cite{Lichard:1993ms,Lichard:1994yx}.

At proton beam experiments $A'$ production is typically taken to be through three main channels: $\pi^0$ decay, $\eta$ decay and proton bremsstrahlung for $m_{A'} \lesssim 2\;\GeV$, see, e.g., \cite{deNiverville:2016rqh,Berlin:2018pwi,Berryman:2019dme}. Putting aside questions about the production kinematics of the parent $\pi^0$ and $\eta$ mesons, their decay kinematics into $A'$ are fixed, and the partial width can be accurately modelled using hadronic form factors determined in ``single Dalitz'' processes $\pi^0, \; \eta \to \gamma (\gamma^*\to e^+e^-)$.\footnote{These form factors are important for the dispersive determinations of $g-2$~\cite{Danilkin:2019mhd}} 
Notably, the form factor measurements are at time-like momentum transfers, as required for decay rate calculations. Repurposing these SM measurements to estimate $A'$ yield provides a simple application of \cref{eq:ap_from_gamma_differential_xsec}~\cite{Batell:2009di}. As an example, the uncertainties in recent determinations of the $\pi^0$ transition form factor~\cite{NA62:2016zfg} are $\sim 15\%$, and even better for $\eta$~\cite{NA60:2016nad}. These are more than adequate for BSM applications at the moment. 

In contrast, models of proton bremsstrahlung require the introduction of form factors evaluated at $q^2$ outside of the observed range~\cite{Foroughi-Abari:2021zbm}. This leads to predictions that depend on unknown hadronic parameters. A common approach to approximate these form factors is to assume they take a dipole form with a single parameter $\Lambda\sim $ GeV, a typical hadronic scale. Uncertainties on observables are then estimated by varying $\Lambda$. Our goal is to avoid these issues by using data and \cref{eq:ap_from_gamma_differential_xsec} to predict dark photon emission rate and kinematics as a function of its mass and coupling.
\subsection{Millicharge Production}
Millicharged particles $\chi$ couple to familiar matter via the electromagnetic interaction:
\beq
\mathcal{L}\supset e Q_\chi A_\mu \bar{\chi}\gamma^\mu \chi.
\label{eq:mcp_coupling}
\eeq
Such particles have been introduced as a component of the dark matter~\cite{Dubovsky:2003yn,McDermott:2010pa,Dvorkin:2019zdi}, to solve experimental anomalies~\cite{Berlin:2018sjs,Munoz:2018pzp}, and as a generic feature of dark sector theories with massless kinetically-mixed vectors~\cite{Holdom:1986eq}. A variety of fixed-target experiments have searched (or whose results have been re-interpreted) for mCPs~\cite{Prinz:1998ua,Magill:2018tbb,ArgoNeuT:2019ckq,SENSEI:2023gie}; many searches using existing or future facilities have been proposed, see, e.g., \cite{Magill:2018tbb,Kelly:2018brz,Gninenko:2018ter,Choi:2020mbk}. Importantly, mCP production inherits the same hadronic uncertainties as the dark photon, because in both cases the underlying process is the emission of an off-shell photon from the target. This makes mCPs a natural second application of the data-driven strategy developed in this work.

Much of the quantitative dark photon discussion also applies to mCPs, except now the parallels to dilepton production are even more straightforward. Specifically, \cref{eq:dilepton_differential_xsec} applies directly with the replacements $m_\ell \to m_\chi$ and $\alpha\to Q_\chi^2 \alpha$. We therefore find the following relation:
\beq
\frac{d\sigma_{\chi\chi}}{dm_{\gamma}^2 d^3 q} 
 =Q_\chi^2 \frac{\sqrt{1-\frac{4m_\chi^2}{m_\gamma^2}}\left(1 + \frac{2m_\chi^2}{m_\gamma^2}\right)}{\sqrt{1-\frac{4m_\ell^2}{m_\gamma^2}}\left(1 + \frac{2m_\ell^2}{m_\gamma^2}\right)} \frac{d\sigma_\gamma}{dm_{\gamma}^2 d^3 q}. \label{eq:mcp_from_gamma_differential_xsec}
\eeq
When applied to production from meson decays, e.g., $\pi^0 \to \gamma \bar\chi \chi$ and $J/\psi\to \bar\chi\chi$, the phase-space integrated version of \cref{eq:mcp_from_gamma_differential_xsec} matches previous results, see, e.g., \cite{Kelly:2018brz}.
\section{Learning From Data}
\label{sec:data_driven}
The main aim of this paper is to provide a recipe for extracting the distributions \cref{eq:ap_from_gamma_differential_xsec,eq:mcp_from_gamma_differential_xsec} from data. Since we do not have access to real event-level data we first construct a mock data set in \cref{sec:data_and_background}.
We will then relate the normalization of the $A'$ or mCP rate (i.e., the total cross-section) to the observed dilepton counts in \cref{sec:production_rate}. Finally, we will construct a fully differential generator for $\Ap$ or mCP kinematics by training a conditional normalizing flow model on the observed kinematic distributions in \cref{sec:kinematics}.

For concreteness we work with NA60 results that provide detailed dimuon spectra, differential in invariant mass and transverse momentum~\cite{NA60:2016nad,NA60:2019tfy}, over a mass range relevant for dark sector searches. This experiment used a 400 GeV proton beam and a composite target. For simplicity we focus on the $L = 0.01$ interaction length lead sub-target of NA60, for which the most information is available.
Similar measurements of $e^+e^-$ pairs were made by NA45~\cite{Agakichiev:1998kip}. In both cases the dilepton spectra were well-fit by ``hadronic cocktail'' models, consisting of two- and three-body decays of various light mesons. Both NA60 and NA45 observed forward dilepton production, with angular acceptances similar to future beam dump searches like SHiP~\cite{Aberle:2839677,Albanese:2878604} and DarkQuest~\cite{Apyan:2022tsd}, and we will assume that these future set-ups will be able to collect similar data sets. DarkQuest is an upgrade of SeaQuest, an experiment already designed to measure forward dimuon production. In contrast, SHiP is being designed with an active muon \emph{veto}~\cite{SHiP:2017wac}, so a dimuon measurement may require a short dedicated run without it.
\subsection{Data and Background}
\label{sec:data_and_background}
\begin{figure*}[t]
\centering
\includegraphics[width=0.45\textwidth]{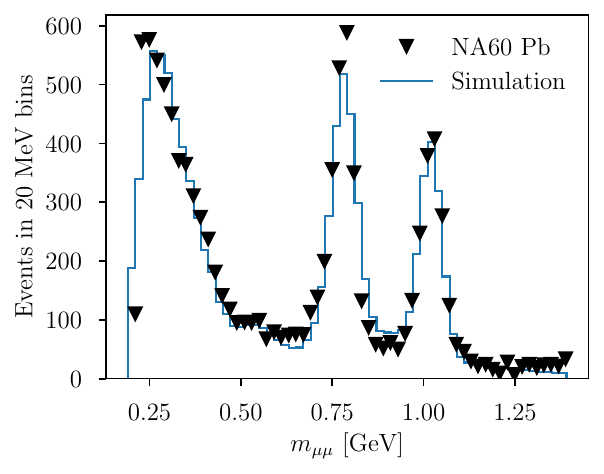}
\includegraphics[width=0.45\textwidth]{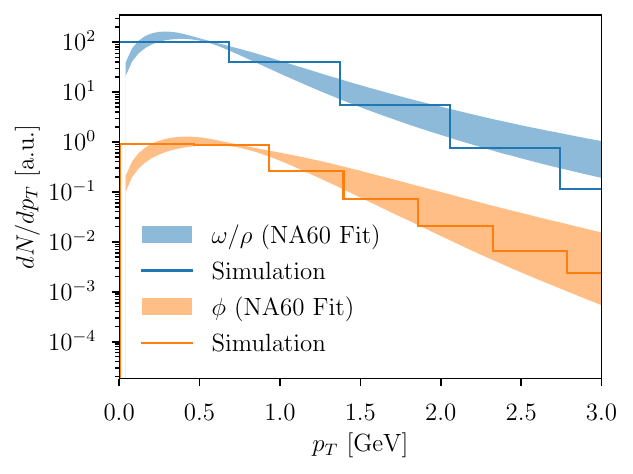}
\caption{Comparison of our tuned mock data set (obtained by reweighing \pythia samples) against dimuon invariant mass (left) and vector meson transverse momentum distributions observed by NA60 from their Pb target. In the right panel, the coloured bands indicate $1\sigma$ ranges of the NA60 $p_T$ spectrum fit parameters.
}
\label{fig:NA60_fit}
\end{figure*}

Our first step is to produce a mock data set using a hadronic cocktail similar to NA60. To do this, we simulate $pp$ collisions in \pythia~\cite{Bierlich:2022pfr} and extract the hadrons used in NA60 fits. 
Specifically, the largest contributions to the dimuon signal come from two-body decays $\eta \to \mu^+\mu^-$, $\rho \to \mu^+\mu^-$, $\omega\to \mu^+\mu^-$, and three-body decays $\eta \to \gamma \mu^+\mu^-$, $\omega\to \pi^0 \mu^+\mu^-$ and $\eta' \to \gamma \mu^+\mu^-$. We require the muons to have at least 3 GeV of energy, an approximate minimum needed for $\mu^\pm$ to pass through the NA60 hadron absorber. 
Each event corresponding to meson $m$ is assigned a weight $w_m$ according to measured branching fraction~\cite{ParticleDataGroup:2024cfk}, total production cross-section $\sigma_m$ and detection efficiency $\mathcal{E}$ reported by NA60~\cite{NA60:2019tfy}:
\beq
w_m = N_{\mathrm{POT}}\br(m \to \mu^+\mu^-X) L \frac{\sigma_m}{\sigma_{\mathrm{inel}}} \mathcal{E}(m_{\ell\ell}^{(m)}, p_T^{(m)}),
\eeq
where $\sigma_{\mathrm{inel}}$ is the total inelastic cross-section in Pb that determines the nuclear interaction length. 
Unfortunately, we do not have complete information to precisely normalize our mock data set: NA60 did not report the exact number of protons on target (POT) collected, $N_{\mathrm{POT}}$, due to a detector malfunction, or the efficiency over the entire dilepton invariant mass range. From \cite{Uras:2011dda} we surmise that $N_{\rm POT} \sim 10^{13}$; the efficiency for three specific meson decays is provided in~\cite{NA60:2019tfy}, which we interpolate to construct an approximate efficiency function $\mathcal{E}(m_{\ell\ell}^{(m)}, p_T^{(m)})$. 

The weights $w_m$ obtained via this procedure are not exact, but provide a rough approximation of the observations; to improve this further, we vary the meson contributions to best fit the published invariant mass spectrum~\cite{NA60:2019tfy}. The resulting weights are different from $w_m$ by $\mathcal{O}(1)$ factors, which is easily attributed to lack of knowledge of the exact POT and acceptance. The resulting fit in terms of invariant mass is shown in the left panel of \cref{fig:NA60_fit}. Note that the observed spectrum is smeared by the $30$ MeV detector resolution~\cite{Uras:2011dda}. As a further check of the kinematics of our mock data, we have reproduced the $m_{\mu\mu}$ lineshapes for the three-body decays given in~\cite{NA60:2016nad}, and the meson $p_T$ distribution (shown in right panel of \cref{fig:NA60_fit}).\footnote{To fully validate the kinematics of our mock data, we would also need to compare dimuon energy distributions (and their correlations with $p_T$) which have not been published.} We therefore conclude that our mock data is a reasonable representation of the experimental results. 

The NA60 dimuon data set includes a combinatorial background contribution at the level of 10\%. These correspond to chance crossings of $\mu^+\mu^-$ tracks from unrelated meson decays that could not be removed through analysis selections. Since we will use event-level data, it will be important to study how the presence of these backgrounds affects the inference of the signal distribution. We therefore also produce a sample of combinatorial background as follows. Using the same \pythia set up as above, we randomly pair opposite sign muons from $\pi^\pm$ and $K^\pm$ decays, weighing each event by the probability of the parent mesons to decay inside the target. 

Open charm production ($\bar{c}c$ with subsequent hadronization into $D$ mesons and their decay into muon-containing final states) also provides a dimuon background, unrelated to $\gamma^*$ production. Such events can be included in the same way as the combinatorial background; however, for simplicity we introduce a $m_{\gamma\gamma} < 1.1\;\GeV$ cut (based on fits in \cite{NA60:2016nad}) to exclude these events when training our normalizing flow models below.  
\subsection{Production Rate}
\label{sec:production_rate}
We now apply \cref{eq:ap_from_gamma_differential_xsec,eq:mcp_from_gamma_differential_xsec} to obtain dark photon/mCP production rates and kinematics.
Experimental selections like angular and energy cuts, detector efficiencies and resolution effects sculpt the fundamental distribution of $\gamma^*$, leading to the shape in \cref{fig:NA60_fit}. These effects can be unfolded on an event-by-event basis~\cite{Andreassen:2019cjw,Canelli:2025ybb}. However, proper unfolding requires a detailed detector simulation which we do not have. To model the result of this unfolding we simply remove the detector acceptance factor from the weights described above. Resolution effects can also be unfolded, but we do not pursue this here for two reasons~\cite{Cowan:1998ji}. Firstly, it would introduce uncertainties that would need to be tracked throughout the analysis. Secondly, resolution smearing smooths out sharp features in the invariant mass distribution, which simplifies training of the normalizing flow model below. Note that we need to exclude the contribution of $\eta\to \mu\mu$ from our data set, since this process is not mediated by a single virtual photon~\cite{Gan:2020aco}. The rate for this process can be extracted from data itself, via the $\eta\to \gamma\mu\mu$ contribution, and so the weight of the $m_{\mu\mu} \approx m_\eta$ bins can be reduced accordingly. Since we have access to the truth-level data, we simply omit the $\eta\to\mu\mu$ events in what follows.

The result of this simple unfolding procedure is a sample drawn from the distribution $d\sigma_{\gamma}/dm_\gamma^2 d^3 q$, which determines the cross-sections and distributions of dark photons and millicharged particles  via \cref{eq:ap_from_gamma_differential_xsec,eq:mcp_from_gamma_differential_xsec}. 
\subsubsection{Dark Photon}
\begin{figure}
\includegraphics[width=0.45\textwidth]{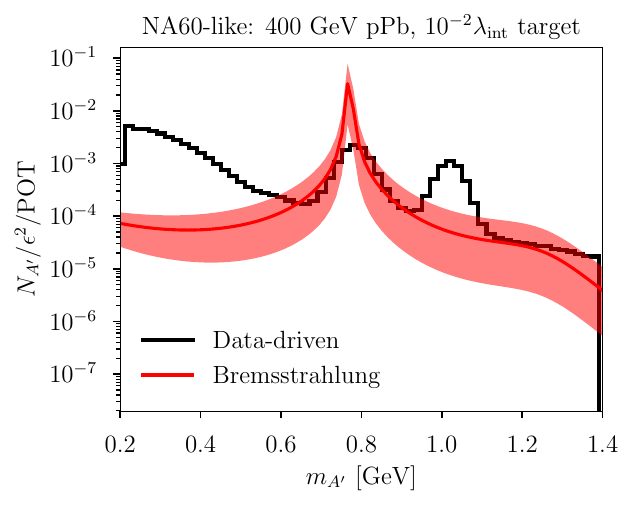}
\caption{Data-driven estimate of $A'$ yield (black line) for a NA60-like beam and target configuration as a function of $m_{A'}$. The mock data includes only meson decays, and not bremsstrahlung (red curve), as the latter contribution was not required to describe the NA60 data~\cite{NA60:2016nad}. The data-driven result accounts for the effects of detector resolution. The red band corresponds to variations of a proton form-factor parameter, in the bremsstrahlung calculation by a factor of 2. Note that both curves correspond to the full phase space without any angular cuts.\label{fig:NA60_yield_comparison}}
\end{figure}

We compute the total $A'$ cross-section by binning our mock data in $m_{\gamma}$ and evaluating $d\sigma_\gamma/dm_\gamma^2$ in each bin using the weights described above. The resulting yield per proton on target is shown in \cref{fig:NA60_yield_comparison}. 
Thus a yield prediction can be made in situ, using the detector and target configuration specific to a given experiment.

Alongside our (mock) data-driven prediction of the $A'$ yield we show a theoretical calculation of proton-nucleus bremsstrahlung $p N \to A' + X$, where $X$ is an arbitrary hadronic final state. This calculation used the \texttt{QRAv2} prescription from \cite{Foroughi-Abari:2024xlj}, form-factor parameters from \cite{Gorbunov:2023jnx}, and we normalized the rate using the non-single diffractive proton-nucleus cross-section given in \cite{Blinov:2021say}. The form-factor used did not include a contribution from the $\phi$ meson, so the corresponding spike in the invariant mass spectrum at $m_{A'} \approx 1\;\GeV$ is absent; more sophisticated models~\cite{Gorbunov:2024vrc,Gorbunov:2024iyu, Kling:2025udr} do feature such an enhancement. 
Note that bremsstrahlung was not one of the processes we used to construct our mock data set, so it is interesting that the yields are comparable in regions where vector meson decays dominate.\footnote{Vector mesons kinematics are generated in \pythia through hadronization of a partonic initial state, rather than coherent bremsstrahlung.} The similarity in rates, however, hides striking differences in detailed kinematic distributions which we will comment on later. Only real data can disentangle different contributions to vector production at these invariant masses. 

The result in \cref{fig:NA60_yield_comparison} is a ``raw'' yield estimate, without any angle cuts (i.e., corresponding to the full phase space of $A'$) or requirements on the detection of the dark photon. For realistic sensitivity predictions one also needs to compute the acceptance for the specific signal to be searched for. In general, this acceptance function is not the same one that determines the SM dimuon yield ($\mathcal{E}$ above). For example, if the goal is to look for displaced decays of the $A'$ the new acceptance function will depend on the location of $A'$ decay vertex, directions of its decay products, etc -- quantities that all depend on the energy and momenta of the emitted $A'$. A realistic calculation of the acceptance is therefore only possible if we can sample $A'$ kinematics at each point of its parameter space (i.e., for different values $m_{A'}$). We show an example of such an acceptance calculation using our trained NFs in \cref{sec:impact_on_acceptance}.

Finally, we note that if the observed distribution $d\sigma_{\gamma}/dm_\gamma^2 d^3 q$ is contaminated by combinatorial backgrounds, the predicted signal yield can be corrected by subtracting the combinatorial yield. This can be inferred from data by studying same-sign lepton pairs, whose production rate is similar to the opposite-sign dileptons that contribute to the background. This is precisely the procedure followed by NA60 to obtain the background-subtracted spectrum shown in \cref{fig:NA60_fit}.
\subsubsection{Millicharged Particles}
\begin{figure}
\includegraphics[width=0.45\textwidth]{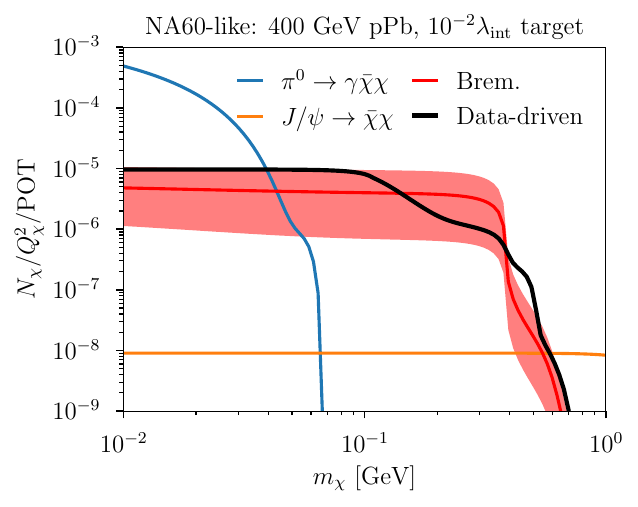}
\caption{Data-driven estimate of mCP ($\chi$) yield (black line) for an NA60-like beam and target configuration as a function of $m_{\chi}$. The mock data includes only meson decays needed to describe the NA60 result and not bremsstrahlung (red curve), which may not account for all relevant production channels in a given mass window. For example, for NA60, contributions from $\pi^0 \to \gamma \bar\chi\chi$ and $J/\psi\to\bar{\chi}\chi$ (shown as coloured lines), are not included due to experimental selections. \label{fig:NA60_mCP_yield_comparison}}
\end{figure}

We obtain the mCP yield by summing the aforementioned weights for events with $m_\gamma > 2m_\chi$, multiplied by the prefactor in \cref{eq:mcp_from_gamma_differential_xsec}. The result is shown in~\cref{fig:NA60_mCP_yield_comparison}. Here we note that experimental selections (such as the focus on dimuon final states, or cuts on $m_{\mu\mu}$) can exclude important production mechanisms, so the (mock) data-driven prediction may not be sufficient. We illustrate this point in \cref{fig:NA60_mCP_yield_comparison} by including production from $\pi^0\to \gamma\bar{\chi}\chi$ and $J/\psi\to \bar{\chi}\chi$, following~\cite{Kelly:2018brz}. These production modes were normalized using $p$Pb $\pi^0$ and $J/\psi$ yields from \cite{Cincheza:1979fy} and ~\cite{NA50:2006rdp,NA60:2010wey}, respectively. Similar to the dark photon case, one can also consider mCP production from hadronic bremsstrahlung $p N \to X + \gamma^*(\bar{\chi}\chi)$. We follow the prescription of \cite{Kling:2025udr} (via \cite{Gninenko:2018ter}) to estimate this yield; this result, shown as the red band in \cref{fig:NA60_mCP_yield_comparison}, is subject to the same hadronic uncertainties as the dark photon case. 
\subsection{Kinematic Distribution}
\label{sec:kinematics}
\subsubsection{Dark Photon}
A measurement of the dilepton invariant mass requires knowledge of the energy $q^0$ and momentum of the virtual photon $\vecbf{q}$. Thus such a data set should provide a fully differential estimate of $d \sigma_{\gamma}/dm^2_\gamma d^3 q$. To make predictions for a dark photon model with a specific $m_{A'}$, however, we are interested in this distribution with the restriction $m_{\gamma} = m_{A'}$. In other words, we want to know the conditional probability density 
\beq
p(\vecbf{q} | m_\gamma = m_{A'}) \propto \left.\frac{d\sigma_\gamma}{dm_{\gamma}^2 d^3 q} \right|_{m_\gamma = m_{\Ap}}.
\label{eq:conditional_distribution}
\eeq
We will further assume azimuthal symmetry to reduce the dimensionality of our distribution; this means that in practice we will work with $q_T$, the transverse momentum (with respect to the beam axis), and $E_\gamma = q^0$, the energy of virtual photon/dark photon, instead of the full three-momentum $\vecbf{q}$.

There are many ways to perform this density estimation. A simple non-parametric approach is to bin the data in all of the kinematic variables. Kinematic samples can be drawn using the accept/reject method on the histogram at the desired $m_{\Ap}$ bin. While simple, such a method suffers from the usual trade-offs between resolution and statistical noise stemming from bin size choice; additionally the underlying smooth distribution is modelled by a piecewise-constant function. More sophisticated methods of density estimation such as kernel density estimation and Gaussian mixture models alleviate these concerns but may have other disadvantages, such as difficulty representing power-law distribution tails (c.f. right panel of \cref{fig:NA60_fit}), inability to model conditional distributions in a simple way, or cumbersome sampling procedure. We avoid these issues by using a normalizing flow (NF)~\cite{kobyzev2020normalizing,papamakarios2021normalizing} to model the conditional distribution in \cref{eq:conditional_distribution}. 

Normalizing flows map simple base distributions (e.g., a multivariate Gaussian) to complex target distributions by applying a series of continuous, invertible and differentiable transformations~\cite{Tabak2010,Tabak2013,dinh2014nice,rezende2015variational}. This enables estimation of complex probability densities, evaluation and efficient sampling of the resulting models. Flow models work by finding a transformation $f$ that relates a data point $\mathbf{x}$ to a latent random variable $\mathbf{z}_0$ which is distributed according to the base distribution. The mapping $f$ is represented by a composition of $K$ invertible transformations $f_i$:
\begin{equation}
    \mathbf{x} = f(\mathbf{z}_0) = \left(f_K \circ \dots \circ f_1\right)(\mathbf{z}_0).
\end{equation}
Each $f_i$ is parameterized by a set of tunable parameters, typically implemented using neural networks. Change-of-variables and chain rule formulas allow one to relate the corresponding data ($p(\mathbf{x})$) and base ($p_0 (\mathbf{z}_0)$) distributions as
\begin{equation}
    \ln p(\mathbf{x}) = \ln p_0 (\mathbf{z}_0) 
    - \sum_{i=1}^{K} \ln \left| \det \left( \frac{\partial \mathbf{z}_i}{\partial \mathbf{z}_{i-1}} \right) \right| .
    \label{eq:nf_ll}
\end{equation}
where $\mathbf{z}_i \equiv (f_i \circ \dots \circ f_1) (\mathbf{z}_0)$. The functions $f_k$ must be carefully chosen to be both bijective, and to enable efficient evaluation of the Jacobian determinants in \cref{eq:nf_ll}. Training the NF model involves maximizing this log-likelihood over the dataset by varying the parameters that specify $f_i$. Once the flow is trained, we can sample $\mathbf{z}_0 \sim p(\mathbf{z}_0)$ and then apply the flow transformation to obtain $\mathbf{x} = f(\mathbf{z}_0)$, a new sample from the model distribution. In conditional NF models, $f_i$ additionally depend on a context variable $\mathbf{c}$, which can be discrete or continuous~\cite{Winkler:2019}. This allows the flow to learn conditional distributions $p(\mathbf{x}|\mathbf{c})$ and sample from the trained model for a given value of $\mathbf{c}$.

Normalizing flows have enjoyed a multitude of applications in particle physics~\cite{ParticleDataGroup:2024cfk,hepmllivingreview}. Closest in spirit to our use case are works that used these models to accelerate Monte Carlo integration/sampling of differential cross-sections or detector simulation~\cite{Stienen:2020gns,Bothmann:2020ywa,Gao:2020vdv,Gao:2020zvv,Heimel:2022wyj,Deutschmann:2024lml,Bothmann:2025lwg,Krause:2021ilc,Krause:2021wez}. There a normalizing flow is trained on samples from a theoretically-known (either analytically or via a simulation) distribution to construct an approximation that is efficient to sample from. We instead want to train a NF on the observed (and unfolded) kinematic distributions to enable the generation of new samples, conditioned on one of the variables. 

In practice we use the following standard architecture for the flow, implemented with the help of \texttt{normflows}~\cite{Stimper2023} and \texttt{PyTorch}~\cite{10.1145/3620665.3640366}. 
We use an Autoregressive Rational Quadratic Spline flow with $K=8$ flow steps, with a linear transformation of the latent variables
between each step~\cite{Durkan2019}. Each flow transformation $f_i$ is a monotonic rational quadratic spline with 6 bins.  
The parameters of these splines are modelled by autoregressive neural networks~\cite{germain2015made} with 2 hidden layers of 32 features each. We assume azimuthal symmetry of the detector, so the flow acts on pairs $\mathbf{x}=(q_T, E_{\gamma})$, with the corresponding invariant masses $m_\gamma$ supplied as the conditioning parameters $\mathbf{c}$. In \cref{sec:nf_training} we show that broad ranges of model and optimizer hyperparameters provide similar performance. For the results presented in the main text, we train on the unweighted mock data set consisting of $\sim 3\times 10^{5}$ events. As expected, the NF performs best in regions with high statistics; performance degrades in sparsely populated mass bins (i.e., high $m_\gamma$). This can be alleviated by weighing the high-$m_\gamma$ data more in training, either using a weighted likelihood~\cite{Stienen:2020gns} or a weighted training data sampler.

Note that our mock data set is larger than the data collected by NA60 (c.f. event counts in the left panel of \cref{fig:NA60_fit}). NA60's POT was relatively modest compared to those targeted by future dark sector searches: for example, DarkQuest aims to collect over $10^{18}$ POT, so its dimuon data set can be significantly larger. Another important difference between NA60 and many dark sector experiments is target thickness: the latter experiments often use thick targets and dilepton production can occur at different positions, which may also be added to the training data set. The distribution of these positions can be important for evaluating the acceptance of long-lived particles. 

\begin{figure}
    \centering
    \includegraphics[width=0.45\textwidth]{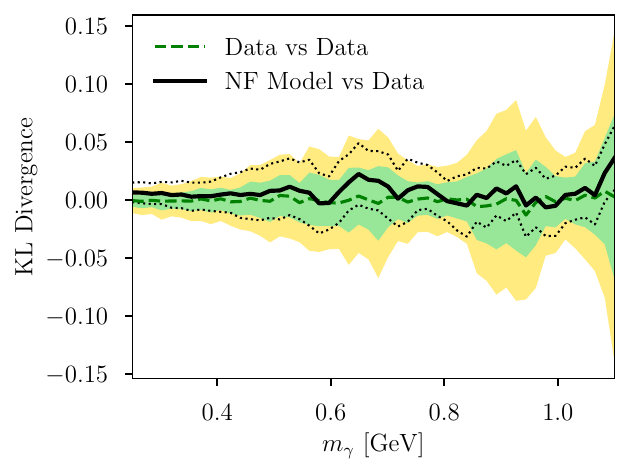}
    \caption{Kullback-Leibler divergence between normalizing flow model trained on mock data and the data itself, as a function of the virtual photon invariant mass $m_{\gamma}$. The green (yellow) bands show 68\% (95\%) confidence limits for KL divergence between different realizations of the data. The black dotted lines indicate the $95\%$ uncertainty bands due to finite sample size drawn from the NF model. The NF model is compatible with the data distribution up to statistical uncertainties, which are more severe at larger invariant masses because of smaller sample sizes.}
    \label{fig:KL_divergence_vs_inv_mass}
\end{figure}

To validate the resulting models we compute the Kullback-Leibler (KL) divergence between the NF model, $p_m$, and the mock data distribution, $p_d$~\cite{burnham2002model,bishop2006pattern}:
\beq
\KL(p_m || p_d) = \int d \mathbf{x}\;  p_m(\mathbf{x}) \ln\left(\frac{p_d(\mathbf{x})}{p_m(\mathbf{x})}\right),
\label{eq:kl_divergence_def}
\eeq
where $\mathbf{x} = (q_T, E_\gamma)$ and both distributions are conditioned on $m_\gamma$.\footnote{The conditional data distribution $p_d$ is approximated using the histogram approach described above, by binning the data in $m_\gamma$ according to the detector resolution.} This quantity vanishes when the two distributions are identical. While the normalizing flow architecture allows for an easy evaluation of $p_m$, we only have samples from $p_d$. We thus estimate $\KL$ using a discrete approximation, following~\cite{wang2006nearest,kahn2024systematic}, with finite samples drawn from the model at different $m_{A'}$ and a subset of the mock data restricted to invariant masses in a small window around $m_{A'}$. This metric is shown in \cref{fig:KL_divergence_vs_inv_mass} as the solid black line. 
In order to understand what constitutes a ``good'' value of this metric, we compare this ``model vs data'' result to $\KL$ between different draws from the mock data set. In the limit of infinite sample size this should vanish. However, depending on the invariant mass region considered, the amount of data varies significantly for a fixed sample size, leading to statistical fluctuations in $\KL$. Reassuringly, the ``data vs data'' divergence (green dashed line) is close to zero across the invariant mass range, while ``model vs data'' is within one standard deviation from this, with larger excursions occurring at values of invariant mass where there were fewer training samples (c.f. \cref{fig:NA60_fit}). In \cref{sec:nf_training} we illustrate how KL improved with larger training sets across the invariant mass range.

\begin{figure*}
    \centering
    \includegraphics[]{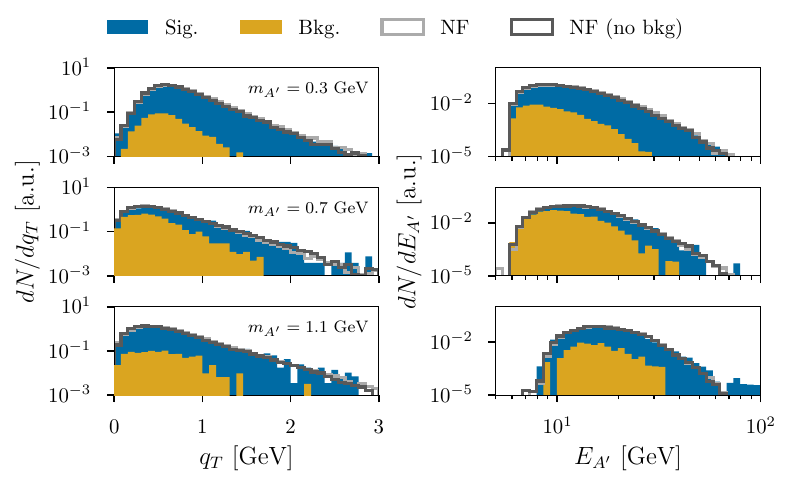}
    \caption{Comparison of the normalized $q_T$ (left column) and $E_{A'}=E_\gamma$ (right column) dark photon distributions sampled from two different NF models and the mock signal dataset for different $m_{A'} = m_\gamma$ values. The filled histograms show the training data (background contribution) in blue (orange) in each invariant mass bin. The model labelled ``NF'' (light gray) was trained on a signal + combinatorial background sample, while ``NF (no bg)'' (dark gray) was trained on a signal-only sample.
    In each row the data is a subset of the full data set with virtual photon invariant masses within 30 MeV of the corresponding $m_{A'}$. Both NF models successfully interpolate the distributions to arbitrary $m_{A'}$. The model trained on the realistic sample that includes the background provides a good description of the signal-only distribution, with differences only becoming apparent in the tails.}
    \label{fig:1d_dist_comparison}
\end{figure*}

As described above, one of the main advantages of normalizing flows for density estimation is the ability to easily sample from the trained model. Because during training $m_{\gamma}$ is provided as a continuous context, the model is able to interpolate between different values of $m_{\gamma}$. In \cref{fig:1d_dist_comparison} we show distributions of kinematic samples drawn from the model at three different values of $m_{A'} = m_\gamma$. While the data set contains a range of $m_\gamma \in [2m_\mu, 1.1\;\GeV]$, the specific $m_{A'}$ values shown in that figure are not present as discrete data points in the training set. However, because the underlying distributions are smooth, they should be close to the observed distribution of events averaged over a narrow range centered on each invariant mass value. We see that this is indeed the case, across the entire relevant invariant mass range. 

We conclude that the NF model can be well-trained with reasonable data set sizes, and offers a faithful approximation to $p(q_T,E_\gamma | m_{A'})$. This enables fast generation of $A'$ kinematics with arbitrary mass (as long as they lie in the range probed by the data), without introducing theoretical models for specific production channels. This data-driven generator can then be used to evaluate efficiency functions for, e.g., long-lived $A'$ searches. 
This quantity, together with the overall $A'$ yield normalization discussed in \cref{sec:production_rate}, can be used to accurately evaluate the sensitivity of such observations. 
\subsubsection{Millicharged Particles}
In contrast to the dark photon case, a wide range $m_\gamma \geq 2m_\chi$ of virtual photon invariant masses contributes to mCP production. Normalizing flows can be used to construct a data-driven MC here as well. Several approaches are possible. The most straightforward is to train a different flow for each $m_\chi$ hypothesis, enforcing $m_\gamma \geq 2m_\chi$ at the training set level.  Another possibility is to use the entire $m_\gamma$ range available in the data, and then perform rejection sampling on $m_\gamma = m_{\chi\chi}$ at generation time. Only one unconditional flow is needed in this case. Finally, one can train the same conditional flow as for the dark photon, and use it in conjunction with sampling from the one-dimensional invariant mass distribution. 
In all of these cases, once a viable $\gamma^*$ sample is generated, it can be easily decayed into $\bar{\chi}\chi$ which can be used for signal acceptance calculations. We have validated the latter approach and obtained comparable agreement to the dark photon case (see \cref{fig:1d_dist_comparison}) between NF-generated and truth-level (i.e., obtained directly from our mock data) $\bar{\chi}\chi$ distributions. 
\section{Discussion}
\label{sec:conclusion}
We have shown that fixed-target dark photon or millicharged particle production can be inferred directly from measured dilepton spectra. The method makes use of a special property of these scenarios: all interactions are electromagnetic in nature. This leads to simple relationships, \cref{eq:ap_from_gamma_differential_xsec} and \cref{eq:mcp_from_gamma_differential_xsec}, between the $A'$ or mCP cross-section and the observed virtual photon distribution, enabling both overall rate predictions and fully differential kinematic sampling without detailed hadronic modeling.

We further demonstrated that conditional normalizing flows can be used to model the observed kinematic distribution. A key feature of these techniques is the ease of sampling from the resulting model. This enables realistic and fast simulation of dark sector signal events, which is necessary to accurately determine signal acceptance. 

A number of questions and possible extensions follow from this strategy. First, the dark photon is only one example of a gauge extension of the SM. Can the same technique be applied to other anomaly-free vector particles, like $B-L$ and $L_i - L_j$ vector bosons? Certainly the situation is more complicated compared to the $A'$ model, since those vector bosons do not just couple to the EM current. However, in certain cases, such a coupling is generated at loop level, as in the $L_\mu-L_\tau$ model. Despite the loop suppression, this induced kinetic mixing is experimentally important~\cite{Blinov:2025aha} and so our method would apply directly to this case. This approach can also be applied to other currents if their SM signatures can be isolated experimentally in analogy to dilepton events. 

\begin{figure}[h!]
    \centering
    \includegraphics[width=0.45\textwidth]{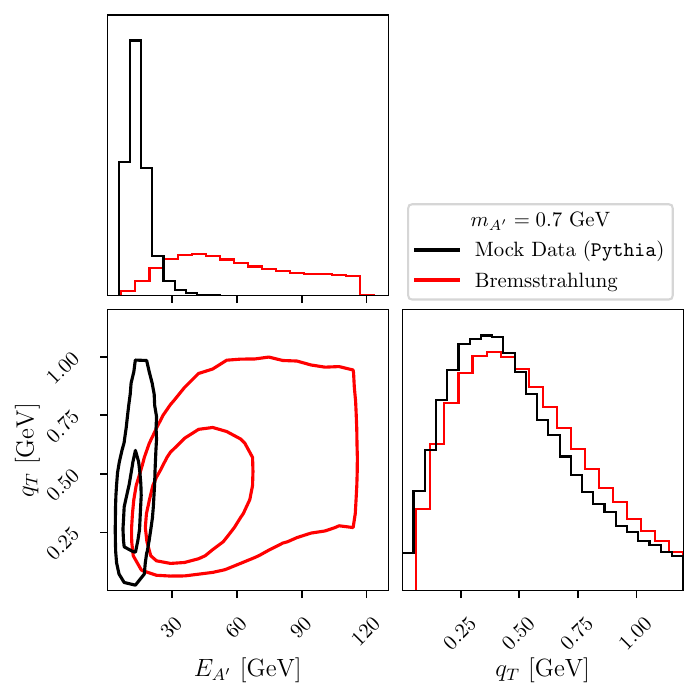}
    \caption{Predicted $A'$ phase space distribution in the $(E_{A'}, q_T)$ plane for $m_{A'} = 0.7$ GeV, sampled from the NF model trained on mock data (signal and combinatorial background, black lines) or from bremsstrahlung (red line). While the $q_T$ distributions are similar, the energy distributions are distinct. The inner (outer) contours indicate $1\sigma$ ($2\sigma$) containment regions.}
    \label{fig:mock_data_vs_brem}
\end{figure}

Another interesting application of fully-differential kinematic distributions would be to disentangle the roles of different production mechanisms of vector mesons and beyond-SM particles. In \cref{fig:NA60_yield_comparison,fig:NA60_mCP_yield_comparison} we illustrated how a data-driven estimate of dark sector particle production can be compared to a theoretical calculation of bremsstrahlung. While there are uncertain inputs (the proton form-factor) to these calculations, the overall rate appears to be consistent with ``data''. However, if we compare the kinematic distributions in \cref{fig:mock_data_vs_brem} we find striking differences (the same observation holds for mCP bremsstrahlung). We emphasize that ultimately \cref{fig:mock_data_vs_brem} is a comparison between \pythia, which produces vector mesons via hadronization, and coherent emission of $A'$ or $\gamma^*(\bar{\chi\chi})$ from a proton. Only real observations can quantify the relative contributions of these mechanisms in different regions of phase space. In particular, \cref{fig:mock_data_vs_brem} suggests that vector energy distributions could be a powerful discriminator between hadronic bremsstrahlung and hadronization for both $\Ap$ and mCP models. Curiously hadronic bremsstrahlung of dilepton pairs also has an uncertain history; it was initially invoked to explain excess dilepton production in high-energy fixed-target experiments at low invariant masses~\cite{Agakichiev:1998kip}. It turned out that the excess was due to underestimated $\eta$ yield, with the bremsstrahlung contribution being highly subdominant~\cite{Akesson:1994mb}. At low, $\sim \GeV$, beam energies, however, hadronic bremsstrahlung is significant, see, e.g.,~\cite{HADES:2009cui}.

Our proposed analysis neglects several important experimental subtleties. First, our desired SM signal, off-shell photons traced by dilepton pairs, is subject to backgrounds. For example, there are combinatorial backgrounds due to chance overlaps of independent muon tracks coming from, e.g., charged meson decays. Much of this background can be eliminated on an event-by-event basis using tracking and tight vertex selections. At NA60, the residual combinatorial background was about 10\% of the total data set~\cite{NA60:2016nad}. After these selections, it is difficult to reject specific events as being combinatorial, but the overall rate of the backgrounds and the shape of their distribution can be estimated from data. Another source of dilepton events are Bethe-Heitler reactions $\gamma Z \to Z \ell^+\ell^-$ and $\eta\to\ell^+\ell^-$, where $\ell^+$ and $\ell^-$ emerge from different vertices in both cases. These events do not contribute to $\gamma^*\to \ell^+ \ell^-$ and therefore should be excluded from the analysis. The contribution of these events to the desired signal can be reduced with kinematic cuts, or by adjusting the weight of the corresponding phase space. The upshot is that realistic event samples are likely not pure samples of virtual photons.  Another source of uncertainty is introduced during the unfolding of dilepton acceptance, which is associated with knowledge of the detector, Monte Carlo or data statistics. All of these ambiguities should translate into uncertainties of the kinematic samples and their relative weights. These can be propagated to the $A'$ model in the Bayesian normalizing flow framework~\cite{trippe2018conditional}; in this architecture the flow parameters are modelled as distributions, rather than fixed numbers. Model uncertainties from these multiple sources can then be assessed by averaging predictions over multiple realizations of the flow parameters~\cite{Bierlich:2023zzd}.

In the future it will be important to assess the utility of our proposed method in the context of actual future facilities, accounting for their target thickness, detector geometry and uncertainties as outlined above.

\begin{acknowledgments}
We thank Yoni Kahn, Shirley Li, Kevin Kelly, Sean Tulin for useful discussions. We are especially grateful to Maxim Pospelov for discussions which initiated this project and for helpful feedback on the manuscript. This work was supported by the Natural Sciences and Engineering Research Council of Canada (NSERC). Our numerical calculations were enabled in part by support provided by Compute Ontario, Calcul Qu{\'e}bec and the Digital Research Alliance of Canada (\href{http://alliancecan.ca}{alliancecan.ca}). This work is made possible by the following open-source software: \texttt{normflows}~\cite{Stimper2023}, \texttt{PyTorch}~\cite{10.1145/3620665.3640366}, \texttt{scikit-learn}~\cite{scikit-learn}, \texttt{corner.py}~\cite{2016JOSS....1...24F}, \texttt{vegas}~\cite{Lepage:2020tgj}, \texttt{numpy}~\cite{harris2020array}, \texttt{scipy}~\cite{2020SciPy-NMeth} and \texttt{matplotlib}~\cite{Hunter:2007}.
\end{acknowledgments}
\onecolumngrid
\appendix

\FloatBarrier 
\section{Normalizing Flow Training}
\label{sec:nf_training}
In this appendix we discuss the details of our normalizing flow model training and hyperparameter selection. Before training, our dataset was log-transformed and standardized by subtracting the mean and dividing by the standard deviation. We trained our model using the \texttt{ADAM} algorithm~\cite{kingma2014adam} for 500 epochs in batches of 1024 points, a learning rate of $10^{-3}$ and an optimizer weight decay of $10^{-5}$. In each epoch $5\times 10^{4}$ data points are randomly drawn from our data set for the model to train on.

Normalizing flow model hyperparameters were chosen by performing a systematic scan over the number of flow steps, hidden layers and features. The mean KL divergence was then computed for each set of hyperparameters over 100 different seed values, so that model-generated data was varied for each KL calculation. KL divergence calculations used $k=2$ nearest neighbours for the density approximation. Since a range of values were found to provide a low mean KL divergence between model-generated points and the data, the final hyperparameter values were chosen to minimize architectural complexity and the number of tunable parameters within the NF transformation. Hyperparameter scans were performed on data without the combinatorial background. All plots in this appendix used data with background contribution.

\begin{figure}
    \centering
    \includegraphics[width=0.45\textwidth]{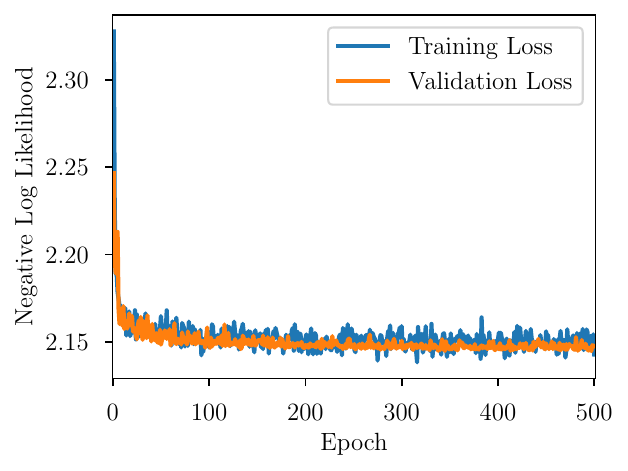}
    \caption{Typical training and validation losses as a function of training epoch for the normalizing flow model.}
    \label{fig:losses}
\end{figure}

The data set was $10^6$ points generated from $10^8$ \pythia events, with $10^5$ points used for plots within the main text, while the full data set was used for appendix figures. The data set was split with 80\% used for training and 20\% for validation. Since past works have shown test data should be at least the same size as training data~\cite{Coccaro_2024}, another set of $10^6$ points was used for testing model results and KL divergence calculations. The training loss was taken as the mean negative log likelihood over all batches of the training data per epoch. The validation loss was calculated once per epoch as the negative log likelihood of the full set of the validation data. A typical set of loss curves as a function of training epoch is shown in \cref{fig:losses}, showing good convergence and absence of overfitting (which would be indicated by an increase in only the validation loss).

\begin{figure}
    \centering
    \includegraphics[width=0.65\textwidth]{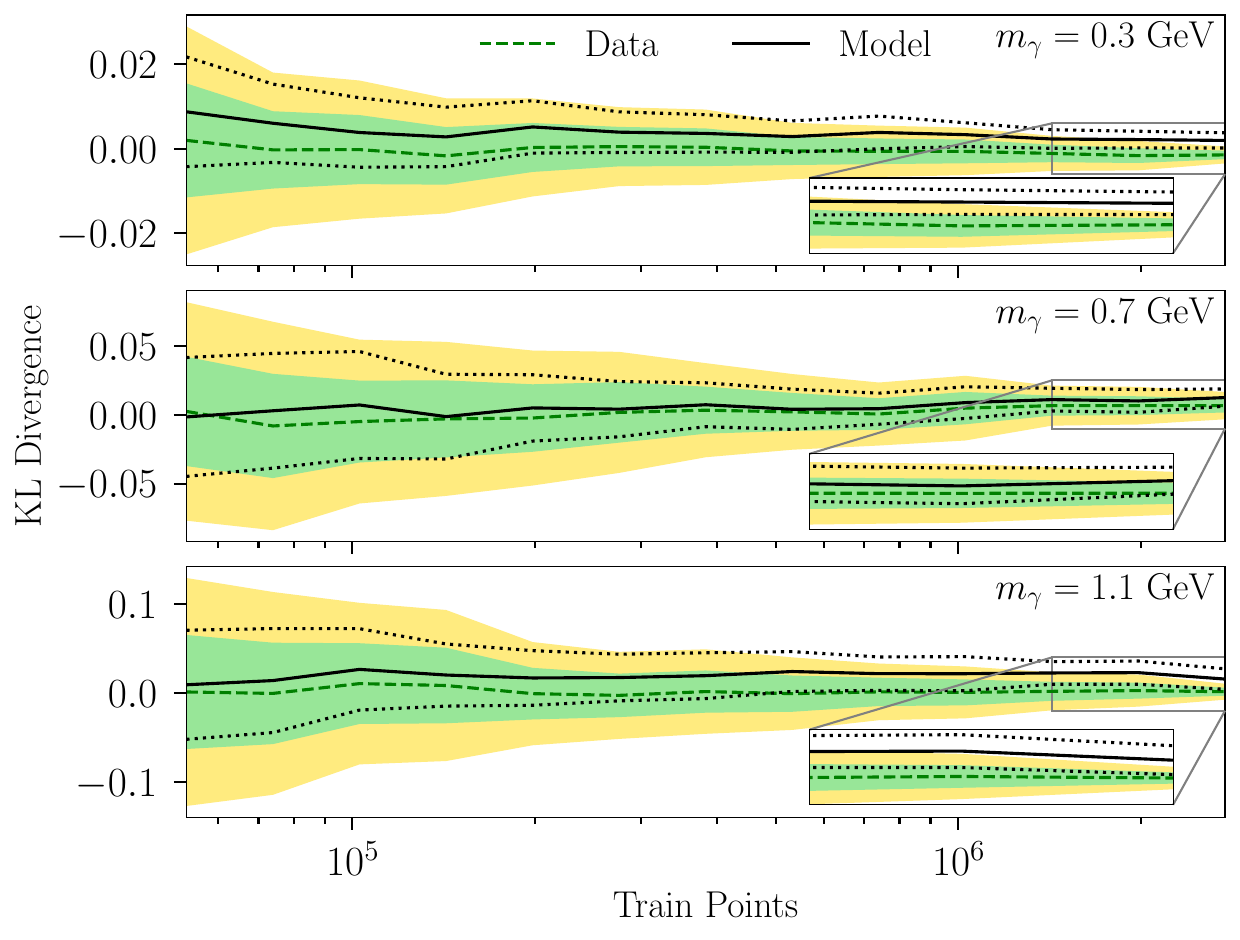}
    \caption{Kullback-Leibler divergence between the normalizing flow model and the data for different $m_{\gamma}$ values as a function of the number of points used to train the model. As in \cref{fig:KL_divergence_vs_inv_mass}, the green (yellow) bands show 68\% (95\%) confidence limits for KL divergence between different realizations of the data. Unlike \cref{fig:KL_divergence_vs_inv_mass}, the black dotted lines now indicate the $1\sigma$ uncertainty bands due to initialization variance, i.e., different initial weights of the NF model prior to training.}
    \label{fig:KL_init}
\end{figure}

Training results in normalizing flow models have been shown to depend on neural network initialization~\cite{kahn2024systematic,Bishop2024}. To quantify this, the model was trained 50 times with random seeds corresponding to different starting model weights. The standard deviation of the mean KL divergence over these runs was 0.002, 0.006 and 0.012 for $m_{\gamma}=0.3$, $0.7$ and $1.1$ GeV respectively, highlighting initialization variance as another source of uncertainty in our model. \cref{fig:KL_init} shows the KL divergence for different $m_{\gamma}$ values as a function of the amount of data used in model training, averaged over different model weight initializations. As the KL divergence between the model and the data is lower for denser $m_{\gamma}$ regions, bias from our approximation method for the mock data distribution becomes more visible~\cite{Zhao2020}, causing the model KL divergence to fall outside of the data KL divergence range. However results still showed that initialization variance decreased as the amount of training points increased. In different architectures, wider networks have been shown to reduce bias~\cite{neal2019}. 
\section{Impact of Combinatorial Background on Acceptance}
\label{sec:impact_on_acceptance}
We have noted that realistic dilepton data sets will include a contribution from events that do not originate from $\gamma^* \to \mu^+\mu^-$ decays. These events arise due to chance overlaps of opposite sign muons from unrelated meson decays, or from ``open-charm'' processes where $c\bar{c}$ production is followed by hadronization and muonic decay of the resulting $D$ mesons. In this appendix we quantify the impact of this data contamination on estimates of dark photon acceptance in a prototypical long-lived particle search. 

\begin{figure}
    \centering
    \includegraphics[width=0.95\textwidth]{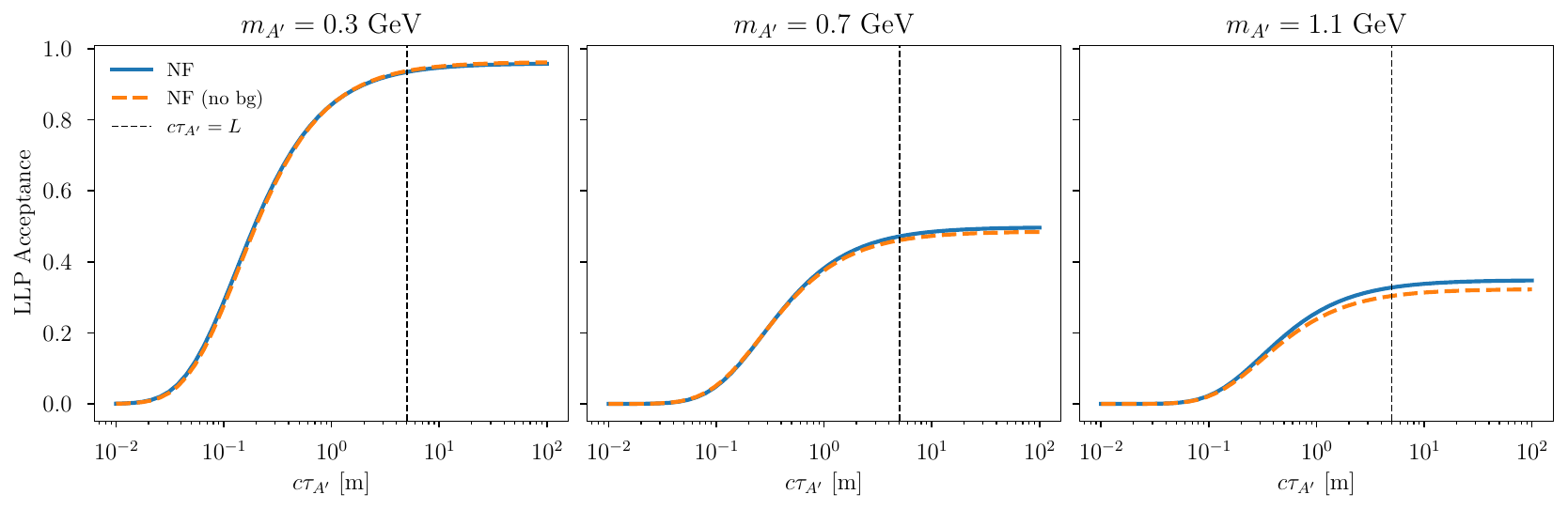}
    \caption{Long-lived particle detector acceptance as a function of decay length for three representative mass points. The acceptance is estimated using \cref{eq:mc_acceptance} for a mock detector described in the text. The solid and dashed lines correspond to $A'$ samples generated using NF models trained on mock data with or without the combinatorial background contribution.}
    \label{fig:llp_acceptance_vs_ctau}
\end{figure}

We consider a mock detector, with geometric acceptance in the range $\theta_{\text{min}} = 35\;\text{mrad}\;< \theta_{A'} < \theta_{\text{max}} = 120\;\text{mrad}$ relative to the beam axis, and a $L = 5$ m shield (motivated by the NA60 set up). We then use our NF models to sample dark photon momenta and compute the acceptance $\mathcal{E}$ 

\beq
\mathcal{E} = \langle  \Theta(\theta_{A'} - \theta_{\text{min}}) \Theta(\theta_{\text{max}}-\theta_{A'})e^{-L / (\gamma v \tau_{A'}) } \rangle
\label{eq:mc_acceptance}
\eeq
where the angle brackets represent an average over the samples. 
We repeat this using NF trained on data with and without combinatorial background. The acceptance for 3 representative mass points is shown in \cref{fig:llp_acceptance_vs_ctau} as function of $c_{\tau_{A'}}$. We observe good qualitative agreement between the two NF models. 

\begin{figure}
    \centering
    \includegraphics[width=0.45\textwidth]{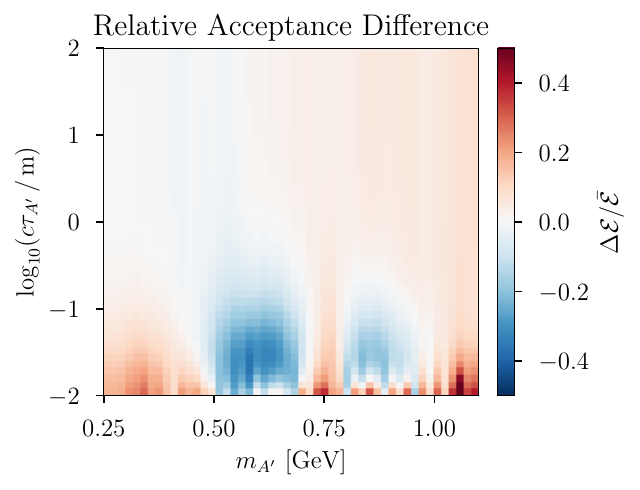}
    \caption{Relative difference in acceptance computed using NF models trained on data with and without combinatorial background in the $m_{A'}$ - $c\tau_{A'}$ plane.}
    \label{fig:delta_acceptance}
\end{figure}

The acceptance in \cref{eq:mc_acceptance} can be exponentially sensitive to the kinematic distribution. We therefore compute the relative difference $2 (\mathcal{E}_{\text{w bkg}}-\mathcal{E}_{\text{no bkg}}) / (\mathcal{E}_{\text{w bkg}}+\mathcal{E}_{\text{no bkg}})$ between background and no-background models across a broad range of parameter space in \cref{fig:llp_acceptance_vs_ctau}. As expected, the acceptance in the long-lived regime is only mildly sensitive to the energy distribution, leading to relative differences of less than 10\% across all masses considered. Shorter lifetimes accentuate the differences between predicted distributions as shown in \cref{fig:delta_acceptance}, but the signal yield is also exponentially falling in that parameter space. An assessment of the impact of this effect would require a more detailed analysis using the total expected yield in a given experiment. Moreover, if a systematic shift in the acceptance can be computed using mock data as illustrated here, it can be corrected for in evaluating the true data-driven acceptance prediction. 

\FloatBarrier 
\bibliographystyle{apsrev4-1}
\bibliography{biblio}
\end{document}